\documentclass[prd,twocolumn]{revtex4}

\usepackage{psfrag}
\usepackage{graphicx}
\usepackage{graphics}
\usepackage{bm}
\usepackage{color}

\newcommand{\be}{\begin{equation}}
\newcommand{\ee}{\end{equation}}
\newcommand{\ba}{\begin{eqnarray}}
\newcommand{\ea}{\end{eqnarray}}

\newcommand{\dcom}[1]{}
\newcommand{\dnote}[1]{}

\newcommand{\gsim}{\raise.3ex\hbox{$>$\kern-.75em\lower1ex\hbox{$\sim$}}}
\newcommand{\lsim}{\raise.3ex\hbox{$<$\kern-.75em\lower1ex\hbox{$\sim$}}}

\begin{document}

\renewcommand{\thefootnote}{\fnsymbol{footnote}}


\renewcommand{\thefootnote}{\arabic{footnote}}
\setcounter{footnote}{0} \typeout{--- Main Text Start ---}

\title{Spontaneous Vortex Production in Driven Condensates with
Narrow Feshbach Resonances}
\author{ Chi-Yong  Lin and Da-Shin  Lee}\affiliation{
Department of Physics, National Dong Hwa University, Hua-Lien,
Taiwan 974, Republic of China}
\author{Ray\ J.\ Rivers}
\affiliation{ Blackett Laboratory, Imperial College\\
London SW7 2BZ, United Kingdom}

\date{\today}
\begin{abstract}

We explore the possibility that, at {\it zero} temperature, vortices can be created spontaneously in a
condensate of cold Fermi atoms,  whose scattering is controlled by a
narrow Feshbach resonance, by rapid magnetic tuning from the BEC to BCS regime.
This could be achievable with current experimental techniques.
\end{abstract}


\maketitle

\section{Causality Bounds}

 Causality imposes strong bounds on a system whose environment is changing rapidly.
 As proposed by Kibble and Zurek \cite{kibble1,zurek1,zurek2} for continuous transitions, where the attempt to achieve infinite correlation lengths in a finite time is inhibited by causality, this can lead to frustration.
 In particular, vortices can arise spontaneously if they can be trapped by domain boundaries.

 Spontaneous vorticity has already been observed
 on condensing cold Bose gases  by rapid cooling \cite{weiler}, supporting the Kibble-Zurek (KZ) scenario.
 In this article we suggest, on similar causal grounds, that vortices can be created spontaneously in a condensate of {\it cold}  {\it Fermi} atoms by rapidly tuning the binding energy of a dominant Feshbach resonance \cite{exp_MolecularBEC} with an external magnetic field.

 Before going into any details we need to think more about the Kibble-Zurek (KZ) scenario which, although originally posed for transitions, does not of itself require a transition, merely a rapid change in correlation length.  In fact, even in temperature quenches across transitions the final defect density is determined by what happens {\it after} the critical temperature is crossed and not what happens before \cite{nuno}. This is demonstrated experimentally  \cite{Monaco2,Monaco1} in the spontaneous production of fluxons on quenching annular Josephson tunneling junctions (JTJs). Here the order parameter is the Josephson phase, and the relevant causal velocity is the Swihart velocity, with no counterpart on the normal conductor side of the transition.  Simulations of idealized JTJs \cite{anna} show how the presence of initial fluctuations just below the critical temperature is sufficient for defects to form with the KZ scaling exponents observed in \cite{Monaco2,Monaco1}. The authors of Ref.\cite{anna} call this mechanism for defect production "fluctuational", rather than the KZ mechanism. We have kept with "KZ scenario" as shorthand, since we always understood the KZ mechanism to work in this way \cite{nuno}.

 This is relevant for the case of magnetic quenches of cold Fermi gases, for which there is no transition, but large and rapidly varying correlation lengths. In such gases weak fermionic pairing gives a BCS theory of Cooper pairs, whereas strong fermionic pairing gives a BEC theory of diatomic molecules.  On driving the condensate from the deep BEC regime toward the BCS regime by ramping an external magnetic field $H$, the speed of sound $v_s$ increases from essentially zero to $O(v_F)$, the Fermi velocity.  However, the (adiabatic) correlation length $\xi$ {\it decreases} as the velocity increases, from a high value when the initial speed of sound is sufficiently small, in accord with the Bogoliubov result $\xi\propto v_s^{-1}$.
 Thus, if the quench is fast enough the condensate has to be frozen initially to prevent the correlation length collapsing acausally fast. Unlike for the case \cite{weiler} above (and JTJs), the effect is observed at $T=0$ (first sound).

An estimate of the time $\bar t$ at which the system unfreezes is \cite{kibble1,zurek2} when it can change no faster i.e.
 \be
 |{\dot\xi}({\bar t})|\approx v_s({\bar t}).
 \label{tbar}
 \ee
 As we shall argue later, provided initial fluctuations are sufficient,  vortices will occur to accommodate the frustration of the field. In the KZ scenario it is suggested that  vortex separation  at their time of spontaneous production is also $O({\bar\xi})$, where ${\bar\xi}\approx\xi (\bar t)$,. That is, there is a one-scale environment in which healing and phase correlation lengths temporarily coincide. If $\xi_0 = k_F^{-1}$, the inverse Fermi momentum which sets the atomic separation scale,  and $\tau_0 = \hbar/\epsilon_F$, the inverse Fermi energy (in units of $\hbar$), we shall show that
 \be
  {\bar\xi}\approx \xi_0(\tau_Q/\tau_0)^{1/2},
  \label{xibar}
 \ee
 provided $\tau_Q\gg \tau_0$.
The timescale $\tau_Q$  is the quench time for the change in the inverse scattering length induced by the changing magnetic field, and is proportional to the quench time $\tau_H$ for the field change. Experimentally, with current techniques, $\tau_Q$ can be made comparable to $\tau_0$ itself, suggesting that spontaneous vortex production should be observable in realistic systems.

In this regard there are many similarities with the analysis of \cite{zurek3} for the thermal quenching of condensates (although in that case causality is determined by second sound), for which the correlation length at unfreezing shows a similar allometric scaling with the quench rate. However, we believe that our approach, in which the condensate remains at $T\approx 0$, has some advantage over the spontaneous production of vortices in thermally quenched
 condensates in that  we have accurate control over the  quench rate of the magnetic field
 in a way that we do not over temperature.   The mechanism that we are invoking here differs from that of defect formation in quantum phase transitions in $T=0$ condensates \cite{zurek4}.

 We stress that our analysis has nothing to do with the
crossover from BEC to BCS regimes which, although characterized by the divergence of
the $s$-wave scattering length $a_S$
\cite{exp_Crossover}, is not a transition.

\section{Cold Fermi gas with a narrow resonance}

For the sake of analytic simplicity, we restrict ourselves to {\it narrow} Feshbach resonances. Our
starting point is the exemplary "two-channel" microscopic action (in units in
which $\hbar = 1$)
 \begin{eqnarray}
S &=& \int dt\,d^3x\bigg\{\sum_{\uparrow , \downarrow}
\psi^*_{\sigma} (x)\ \left[ i \
\partial_t + \frac{\nabla^2}{2m} + \mu \right] \ \psi_{\sigma} (x)
\nonumber \\
   &+& \phi^{*}(x) \ \left[ i  \ \partial_t + \frac{\nabla^2}{2M} + 2 \mu -
\nu \right] \ \phi(x) \nonumber \\
&-& g \left[ \phi^{*}(x) \ \psi_{\downarrow} (x) \ \psi_{\uparrow}
(x) + \phi(x)  \psi^{*}_{\uparrow} (x) \ \psi^{*}_{\downarrow} (x)
\right]\bigg\} \label{Lin}
\end{eqnarray}
for cold ($T = 0$) fermionic fields $\psi_{\sigma}$
 with spin label $\sigma = (\uparrow, \downarrow)$, which possess a {\it narrow} bound-state
 (Feshbach) resonance with tunable
binding energy $\nu$, represented by a diatomic field $\phi$ with
mass $M =2m$. This model has been discussed on great detail by Gurarie and Radzihovsky \cite{gurarie} and we borrow several results from their paper.

 ${S}$ is quadratic in the Fermi fields. Integrating them out \cite{rivers} enables
 us to write ${S}$ in the non-local form
  \begin{eqnarray}
 S_{\rm NL} &=& -i\,{\rm Tr}\ln {\cal G}^{-1}
 \nonumber
 \\
 &+& \int dt~d^3x\,\phi^{*}(x) \ \left[ i  \ \partial_t + \frac{\nabla^2}{2M} + 2 \mu -
\nu \right] \ \phi(x),
\label{SNL}
 \end{eqnarray}
 in which
  ${\cal G}^{-1}$ is the
inverse Nambu Green function,
\begin{eqnarray}
 {\cal G}^{-1} &=& \left( \begin{array}{cc}
        i \partial_t - \varepsilon         &-g\phi(x) \\
                  -g\phi^{*}(x) & i \partial_t +
                  \varepsilon
                  \end{array} \right)
                  \end{eqnarray}
 where
 $ - g\,\phi (x) = g|\phi(x)| \ e^{i
\theta (x)}$
  represents the  condensate (and
$\varepsilon = - \nabla^2/2m - \mu $).

In this paper we restrict ourselves to the {\it mean-field
approximation}, the general solution to $\delta S_{\rm NL} =0$, valid
if $\phi$ is a sufficiently narrow resonance \cite{gurarie,dieh}.
 The action
possesses a $U(1)$ invariance under
$\theta\rightarrow\theta +\rm{const.}$,
 which
is spontaneously broken.
 $\delta\,S_{\rm NL} = 0$ permits the spacetime constant {\it gap} solution $|\phi (x)| = |\phi_0|\neq
 0$.  We perturb in the derivatives of $\theta$ and the {\it small}
fluctuation in the condensate density
 $\delta |\phi|= |\phi|  -  |\phi_{0}|$ and its derivatives, most conveniently in powers of the Galilean-invariant $\Sigma$, defined by
 \begin{equation}
   {\cal G}^{-1} =  {\cal G}_0^{-1} - \Sigma,
 \end{equation}
where
\begin{eqnarray}
 {\cal G}_0^{-1}
        &\equiv &\left( \begin{array}{cc}
        i \partial_t - \varepsilon         &g|\phi_0| \\
                  g|\phi_0| & i \partial_t +
                  \varepsilon
                  \end{array} \right)  \label{greenfun}
                  \end{eqnarray}
$\theta (x)$ is not small. Using the results of our earlier papers \cite{rivers,rivers2}, for which (\ref{Lin}) is a limiting case, at {\it second} order in $\Sigma$
we can extract from $S_{NL}$
 a {\it local} effective Lagrangian
density
      \begin{eqnarray}
 L_{\rm eff} &=& -\frac{1}{2}{\rho}_0
   G(\theta, {\epsilon})  + \frac{N_0}{4}\ G^2(\theta, {\epsilon})
 \nonumber\\
 && -{\alpha}{\epsilon}G(\theta, {\epsilon})
 +\frac{1}{4}{\eta}X^2({\epsilon},\theta)
  -\frac{1}{4}{\bar M}^2{\epsilon}^2,
 \label{LeffU0}
 \end{eqnarray}
valid for long wavelength, low-frequency phenomena.

$L_{\rm eff}$ is given in terms of the Galilean
 scalar combinations
$G(\theta, \epsilon ) = \dot{\theta} + (\nabla
\theta )^2/4m + (\nabla \epsilon )^2/4m$, $X(\epsilon,\theta ) =
\dot{\epsilon}+ \nabla \theta .\nabla \epsilon/2m$,
where the dimensionless $\epsilon\propto\delta
 |\phi|$ is itself a scalar. In particular,
  $\rho_0 = \rho^F_0 +
\rho^B_0 $ is
 the total (fixed) fermion number density where
$\rho^F_0$  is the explicit fermion density,
$$\rho^F_0 = \int d^3 {\bf p} / (2\pi)^3 \ \left[ 1 -
\varepsilon_{p}/E_{p}
  \right] $$
  and $\rho^B_0 = 2|\phi_{0}|^2$ is
due to molecules (two fermions per molecule).  In conventional notation $\epsilon_{k}=k^2/2m$ and $E_{p}=(
\varepsilon_{p}^2 + g^2|\phi_{0}|^2 )^{1/2} $.
For the evolving system the molecular density is
$\rho^B = 2|\phi|^2 = \rho^B_0 + 4\delta |\phi||\phi_0|$, showing that $\epsilon\propto \delta\rho^B$, the molecular density fluctuation.
Although the details are immaterial, we have scaled $\epsilon$ so that it has the
 same coefficients as $\theta$ in its spatial derivatives.

\subsection{The speed of sound}

For small fluctuations, the linear approximation to the Euler-Lagrange equations for $\theta$ and $\epsilon$,
sufficient to determine the speed of sound, is
\ba
 \frac{N_0}{2}\ddot\theta -\frac{\rho_0}{4m}\nabla^2\theta - {\alpha}{\dot{\epsilon}} &=& 0
 \nonumber
 \\
 \frac{\eta}{2}\ddot{\epsilon} -\frac{\rho_0}{4m}\nabla^2{\epsilon}+ \frac{1}{2}{\bar M}^2{\epsilon} + {\alpha}{\dot{\theta}} &=& 0.
 \label{modes}
\ea
 On diagonalizing, we see that for long wavelengths the gapless phonon has dispersion relation
 $\omega^2 = v_s^2 {\bf k}^2$,
 with speed of sound
 \be
  v_s^2 = \frac{\rho_0/2m}{N_0 + 4{\alpha}^2/{\bar M}^2},
  \label{v-s}
 \ee
independent of $\eta$. After renormalisation \cite{rivers} the coefficients in (\ref{v-s}) are
\begin{eqnarray}
N_0 &=& g^2 |\phi_0|^2\int \frac{d^{3} {\bf p}}{(2\pi)^3} \frac{1}{ 2 E_{p}^3} \, ,
\\
\alpha&=& 2|\phi_{0}|+ g^2|\phi_0| \int \frac{d^3 {\bf p}}{(2\pi)^3}
\frac{\varepsilon_{p}}{ 2E_{p}^3} \, ,
\\{\bar M}^2 &=& 4(\nu- 2\mu)-2 g^2 \int \frac{d^3 {\bf p}}{
(2\pi)^3} \bigg[\frac{\varepsilon_{p}^2}{ E_{p}^3} - \frac{1}{ ({\bf
p}^2/2m)}\bigg] \, .
\label{defs}
\end{eqnarray}
The $s$-wave scattering length $a_S$ is determined from the binding energy as
 \be
 2\mu - \nu = \frac{g^2 m}{4\pi a_S}.
 \label{Hdep0}
\ee
To see the effect of applying an external magnetic field $H$, we adopt the parametrisation
  \cite{gurarie}
\be
 a_S = a_{\rm bg}\bigg(1 - \frac{H_{\omega}}{H- H_0} \bigg),
 \label{aSH}
\ee
whence
 \be
 2\mu - \nu \approx -\frac{g^2m}{4\pi a_{bg}H_{\omega}}(H - H_0).
 \label{Hdep}
\ee
 In (\ref{Hdep}) $a_{\rm bg}$ is the background (off-resonance) scattering length, $H_{\omega}$ the so-called "resonance width" \cite{gurarie} and $H_0$ is the field required to achieve infinite scattering length (the unitary limit).  As $H$ {\it increases} through $H_0$ we pass from the BEC to the BCS regimes (with $a_S$ going from +ve to -ve).

As a result, in the deep BCS regime (small $\alpha^2/{\bar M}^2$) $v_s\rightarrow v_{BCS} = v_F/\sqrt{3}$ and in the deep BEC
regime (large $\alpha^2/{\bar M}^2$) $v_s\rightarrow 0$.

An exemplary graph of $v_s^2$ is given in Fig.1.
To a fair approximation, $v_s^2/v_F^2$ can be approximated as
\be
 v_s^2\approx (v_F^2/6)[1+ \tanh (c_0 - b_0/k_Fa_S)].
 \label{tanh}
\ee

 \begin{figure}
\centering
\includegraphics[width=\columnwidth]{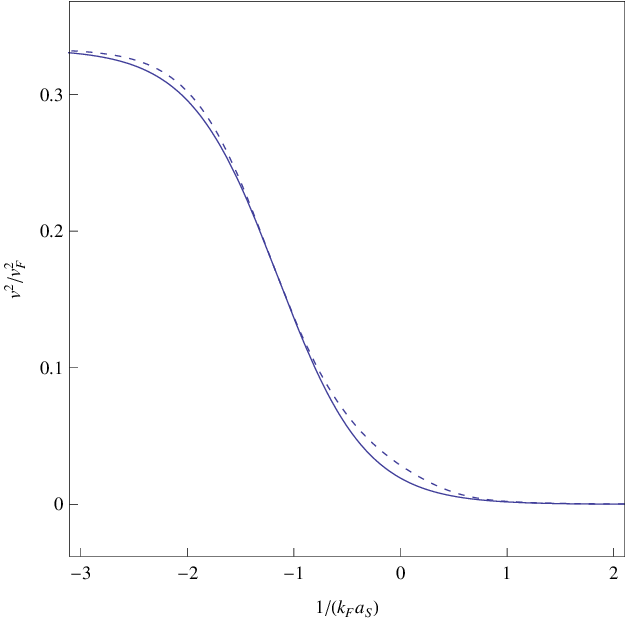}
 \caption{The dotted line shows $v_s^2$ for the value ${\bar g} = 0.9$ as a function of $1/k_Fa_S$. The solid line shows the parametrisation (\ref{tanh}) for $c_0 = -1.398$ and $b_0 = 0.202$. We get as good or better fits for other values of $\bar g$, with $b_0(g)$ varying by only 25\% over the range $0.2 \leq {\bar g}\leq 1.6$. } \label{Fig1}
\end{figure}

 \section{The hydrodynamic approximation and the GP equation}

Equations (\ref{modes}) represent a two-component system of molecules and atom pairs, but
   a two-component density in which fermions  oscillate from one to the other is not the same as a two-fluid picture.   In general, it is a {\it no}-fluid picture. Density fluctuations act as sources and sinks in the continuity equation and the condensate behaves as a fluid only when these are ignorable i.e. we can neglect the spatial and temporal variation  of $\epsilon$, in comparison to $\epsilon$ itself.
 In that case, ${\epsilon}\approx
 - 2{\alpha} G(\theta)/{\bar M}^2$, a slave to the phase,
 whence
    the Euler-Lagrange equation for $\theta$ is, indeed, the continuity equation of a {\it single} fluid,
\be
  \frac{\partial}{\partial t}\rho +\nabla \cdot (\rho{\bf
      v}) = 0,
      \label{contU0}
  \ee
 with $\rho =\rho_{0} + 2{\alpha}{\epsilon} - N_0 G(\theta )$
and ${\bf v} = \nabla\theta/2m$.
To complete the fluid picture we observe that this definition of $\rho$ is no more than the
Bernoulli equation
 \begin{equation}
 m\dot{\bf v} + \nabla \bigg[\delta h + \frac{1}{2}m\,{v}^2 \bigg] =
 0,
 \label{BernU0}
\end{equation}
where the enthalpy
 $\delta h = m v_s^2\delta\rho /\rho.$
The resulting equation of state is $dp/d\rho = mv_s^2$  across the whole regime.

The hydrodynamic equations can be derived from a Gross-Pitaevskii  (GP) equation on ignoring quantum pressure. It is more transparent to reconstitute this equation, with its natural vortex solutions, than to work with the Euler-Lagrange equations for $\theta$ and $\epsilon$ directly.

Consider the Lagrangian describing the wave-function $\psi$ of a particle of mass $2m$, interacting non-linearly with itself,
 \be
L({\psi}) = i\hbar\psi^{\ast}\dot{\psi} - \frac{\hbar^2}{4m}\nabla\psi^{*}
\cdot\nabla\psi - \frac{mv_s^2}{\rho_0}(|\psi|^2 - \rho_0)^2,
\label{GP}
\ee
where we have restored factors of $\hbar$.
The Gross-Pitaevskii equation
following from (\ref{GP}) is
 \be
 i\hbar\dot{\psi} + \frac{\hbar^2}{2m}\nabla^2\psi + 2mv_s^2\psi - \frac{2mv_s^2}{\rho_0}\psi|\psi|^2 = 0,
  \label{GP2}
 \ee
 with coefficients varying smoothly as we cross the unitary limit.
 If we set $\psi = \sqrt{\rho}\;\exp(i\theta)$
and solve (\ref{GP2}) at the relevant order in derivatives, we recover
(\ref{contU0}) and (\ref{BernU0}) on ignoring $\epsilon$ density fluctuations (see \cite{aitchison} for a comparable analysis).
Remarkably, the dependence of the GP equation on the coefficients of (\ref{LeffU0}) is implicit, through $v_s^2$. The length scale of the system is
 $\xi = \hbar/2mv_s = \hbar/Mv_s$
 and the time scale is $\tau = \hbar/Mv_s^2$.

 We stress the importance of the narrowness of the resonance for the single-fluid approximation. At best, even in the hydrodynamic limit broad resonances  require a two-fluid description \cite{rivers,rivers2,aitchison} from which we have no simple way to draw conclusions about defect formation in a field quench.

\section{When is the GP equation valid?}

For narrow resonances the single-fluid approximation given above, with its attendant GP equation, is not valid everywhere, as follows from (\ref{modes}).
Consider a spatially {\it homogeneous} condensate. Then these linearized EL equations, which ignore damping,  display the oscillatory solution
 \be
\delta\rho^B = \delta\rho^B_0 \cos\Omega t,
\label{Omega}
 \ee
 describing the repeated dissociation of molecules into atom pairs and their reconversion into molecules.
 The frequency $\Omega$ of density fluctuations is determined by the energy scale at the beginning
 of the two-fermion cuts in the energy plane arising from the integration over fermionic modes leading to (\ref{SNL}) ($E_{\rm min}$ of
 \cite{gurarie2}) and not
  that of the gapped mode present in the formalism (and of which we need nothing in the context of this paper).
 It can be shown  \cite{gurarie2,timmermans} to increase monotonically from the exponentially damped
 $g|\phi_0| = O(\mu\exp(-\pi/2 k_F|a_S|)$ in the BCS regime to $\Omega = 2\sqrt{g^2|\phi_0|^2 + \mu^2}$ as  it crosses into the BEC regime.
 In the deep BEC regime $\Omega\approx 2\mu$.

The single-fluid approximation requires that density fluctuations can be averaged to zero on the timescale $\tau$ i.e.
$$\tau\Omega \gg 1.$$ Whether this can be achieved or not depends upon the quench.

Suppose that $H$ increases uniformly in time with ${\dot H}/H|_{H_0} = {\tau_H}^{-1}$,
where $t=0$ is the time at which the system is at the unitary limit of infinite scattering length. We write $1/k_Fa_S(t) = -t/\tau_Q$ for small $t$, where
  \be
  \tau_Q = \tau_H\bigg(\frac{k_Fa_{\rm bg}H_{\omega}}{H_0} \bigg).
  \label{tauQ}
  \ee
The quench parameters are related to the width of the resonance  $\Gamma_0$   by \cite{gurarie}
 $\Gamma_0\approx 4m\mu_B^2a_{\rm bg}^2H_{\omega}^2/\hbar^2$,
where $\mu_B$ is the Bohr magneton. In practice, it is more convenient to work with the dimensionless width
 $\gamma_0\approx\sqrt{\Gamma_0/\epsilon_F}$, whereby
  \be
 \frac{\tau_Q}{\tau_0} = \frac{\tau_Q\epsilon_F}{\hbar}\approx \frac{\pi}{\mu_B{\dot H}}\frac{\epsilon_F^{2}}{\hbar}\gamma_0.
 \label{tauQ}
 \ee
To be concrete, consider the narrow resonance in $^6{\rm Li}$ at $H_0 = 543.25 {\rm G}$, discussed in some detail in \cite{strecker}. [This is to be distinguished from the very broad Feshbach resonance in $^6{\rm Li}$ at $850~{\rm G}$.] As our benchmark we take the achievable number density $\rho_0\approx  3 \times 10^{12} cm^{-3}$, whence $\epsilon_F\approx  7 \times 10^{-11} eV$ and $\gamma_0\approx 0.2$. In terms of the dimensionless coupling $\bar g$,
 where $g^2 = (64\epsilon^2_F/3 k_F^3){\bar g}^2$, $^6Li$ at the density above corresponds to ${\bar g}^2\lesssim 1$.
 For a condensate of density $\rho$ it follows that ${\bar g}^2\approx (\rho_0/\rho)^{1/3}$ and
 \be
 \frac{\tau_Q}{\tau_0}\approx \frac{1}{{\dot H}}\bigg(\frac{\rho}{\rho_0}\bigg),
 \ee
 where $\dot H$ is measured in units of $Gauss\, (ms)^{-1}$.  Experimentally, it is possible to achieve quench rates as fast as ${\dot H}\approx 0.1 G/ms$~\cite{strecker}.

 The condition $\tau\Omega\gg1$  throughout the quench now becomes
 \be
 \tau\Omega\ = \frac{\Omega}{\epsilon_F}\frac{v_F^2}{4v_s^2}\approx\frac{\Omega}{\epsilon_F}\frac{\tau_Q}{\tau_0}\gg 1.
 \label{singlefluid}
 \ee
 In the BEC regime, with large $\Omega\approx 2|\mu|$,
 the inequality is guaranteed for all $\tau_Q/\tau_0\gg 1$. However, in the BCS regime $\Omega$ is exponentially damped, violating the inequality (\ref{singlefluid}). Thus the single fluid approximation, on which our causal analysis depends, requires that the quench does not trespass beyond the unitarity limit in the BCS direction before the system unfreezes. The question is then whether such a quench can be implemented sufficiently fast for this to be the case.

\section{Spontaneous vorticity}

There are several length scales in the theory, not all visible in the GP action, which has disconnected the gapless sector (Goldstone sector) from the gapped (Higgs) mode. Our correlation length $\xi$ is the healing length for the fermion density $\rho$ but, as we have observed, in the KZ scenario this is equated to the {\it phase} correlation length at the moment at which vortices are formed spontaneously.

Beginning in the deep BEC regime, where $\xi$ is as large as the system, we need phase fluctuations to seed vortices (as for Josephson junctions \cite{anna}). Normally, treated in isolation,
the GP action (\ref{GP}) and GP equation (\ref{GP2}) would not be a helpful starting point since the chemical potential term in the GP action (the quadratic term in $\psi$) can be eliminated, or otherwise modified, by a change of phase linear in time. However, $\dot\theta$ is now pinned to a small $\epsilon$, prohibiting this, and we can use the potential in (\ref{GP}) to estimate fluctuations. It follows that (with a critical temperature $T_c\sim mv_F^2$), at speed of sound $v_s$ thermal fluctuations at a temperature $T\sim (v_s^2/v_F^2)T_c$ are sufficient to give large fluctuations in the phase. With $v_s$ extremely small in the BEC regime, initially even small thermal fluctuations will be sufficient to permit vortex creation.

We now ramp the magnetic field (constant $\dot H$) to drive the gas from the deep BEC regime to the BCS regime, as in the previous section. Remembering that $t=0$ is the time at which the scattering length $a_S$ diverges,
the time $\bar t$ at which the system unfreezes [as defined by (\ref{tbar})] satisfies
\be
 [v_s^2(t)]^2\approx \frac{\hbar}{2M}\frac{d}{dt}v_s^2(t).
 \label{vbar}
\ee
Our narrow resonance expression (\ref{v-s}) for the sound speed in the BEC regime reproduces  that of \cite{gurarie}, obtained there by different means.
In Fig.2 we have shown where the left- and right-hand sides of (\ref{vbar}) intersect as a function of $t$ or, equivalently, $1/k_Fa_S(t)$.

 \begin{figure}
\centering
\includegraphics[width=\columnwidth]{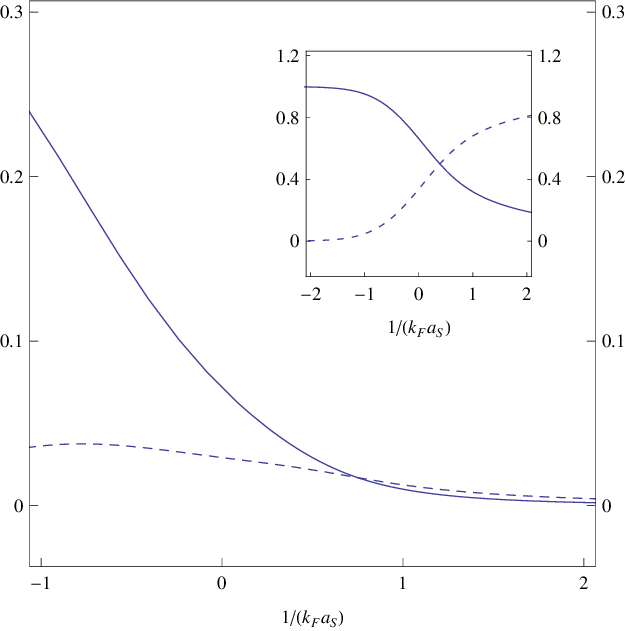}
 \caption{A condensate with ${\bar g} = 1.5$, quenched by a magnetic field with quench rate $\tau_Q/\tau_0 = 16$. For the solid line the ordinate is $v_s^2/v_F^2$ and for the dashed line the ordinate is $\sqrt{\hbar/2M (d/dt) v_s^2}$. In driving the system from the BEC regime (on the right) towards the BCS regime (on the left) the adiabatic regime lies to the left of $1/k_Fa_S\approx 0.75$, when the two curves intersect, 
 from (\ref{vbar}). The inset shows $\rho_0^B/\rho_0$ (dashed curve) and $\rho_0^F/\rho_0$ (solid curve). } \label{Fig1}
\end{figure}
Empirically, we find the approximate universal behavior
\be
v_s^2({\bar t})\approx  v_{F}^2(\tau_0/4\tau_Q)
\label{tbar2}
\ee
for a very wide range of parameters. This universality is not surprising. If we adopt the parametrisation (\ref{tanh}), $v_s^2/v_F^2\propto \tau_0/\tau_Q$ automatically.
The correlation length at the time of unfreezing then satisfies the scaling law (\ref{xibar})
\be
  {\bar\xi}\approx \xi_0(\tau_Q/\tau_0)^{1/2}
  \label{xibar2}
 \ee
 provided $\tau_Q\gg \tau_0$.

Equation(\ref{GP2}) permits vortex creation in which ${\bar\xi}$ determines vortex width, both in fermion number density and phase correlation length. We can ignore the order parameter fluctuations $\delta|\phi|$, since they are shorter-ranged and do not affect how vortices pack. The KZ picture then gives an estimated vortex separation at time of production as (\ref{xibar2}), as anticipated in the introduction.

  We believe it possible to produce vortices spontaneously for realistic quenches. It is simplest to adopt a density $\rho <\rho_0$, permitting smaller $\dot H$ to maintain $\tau_Q > \tau_0$. As an example see Fig.2, the tail of the velocity profile, in which we take ${\bar g} = 1.5$, corresponding to $\rho\approx 0.1\rho_0$, and $\tau_Q/\tau_0 = 16$. Then, the domain structure is formed when $v_s\approx v_{BCS}/4$, with $1/k_Fa_S\approx 0.75$ and $\rho_0^F\approx 0.38\rho_0,  \rho_0^B\approx 0.62\rho_0$. $\tau\Omega$ decreases throughout the quench, but its final value of $16\Omega/\epsilon_F\approx 27$ remains sufficiently large to justify the approximation.  Further, explicit calculation shows that spontaneous vorticity arises only when $|a_S| > r_0$, the range of the interaction, reinforcing the validity of our approximation. Adjacent parameter values are equally successful. Finally, we see that the approximation from Eq.(\ref{aSH}) to Eq.(\ref{Hdep}) is justified, with a fractional error less than
 $k_Fa_{bg}\ll 2. 10^{-2}$.

For the case in hand, with ${\bar \xi}\approx 4\xi_0$, spontaneous vortex creation should be possible, since
 the
 length scale $\xi_{c}$ for a condensate of $N = 10^5$ atoms at this density
 would give $\xi_{c}\approx 100 \xi_0$. For pancake traps the transverse width of the condensate is even larger. This suggests that a large number of vortices should be created, but some caution is required. What is not usually stressed is that the KZ prediction (\ref{xibar}) is, actually, and {\it upper} bound. The extent to which this bound is saturated depends on the system. To cite extremes for spontaneous vortex formation at thermal quenches, it is saturated for vortex production on quenching $^3He-B$ \cite{helsinki}, but underestimates vortex separation strongly for high-$T_c$ superconductors \cite{technion}. The spontaneous creation of vortices in thermal quenches on low-$T_c$ superconductors \cite{Monaco2} and fluxons in Josephson junctions \cite{Monaco1} give results in between.

\section{Outlook}
 We believe that spontaneous vortex creation by a magnetic field ramp at narrow resonances such as the $^6Li$ resonance at $543G$ could be achievable with current experimental techniques. This follows from our single-fluid approximation and its concomitant Gross-Pitaevskii equation, valid because the vortices form sufficiently early in the ramp that we do not have to continue into the BCS regime where it fails.

There is the further advantage that, although our idealized
calculations were for temperature T = 0, in reality temperature is finite. By stopping
soon enough, we would hope to remain clear of critical thermal behavior since we are going from a regime of negative chemical potential to one of positive chemical potential. This has been discussed in the final paragraph of \cite{strecker}. Furthermore, even very small thermal fluctuations are sufficient to kickstart defect production.

 Narrow resonances are difficult to work with because of the required field stability, but we expect them to give most defects after a ramp.
 Increasing resonance width in (\ref{tauQ}) increases $\tau_Q$ and hence $\bar\xi$ at fixed density. However, with
 ${\bar\xi}\propto \gamma_0^{1/2}$ for moderately narrow resonances, the effect of broadening the resonance is, initially, weak and we can still anticipate observable spontaneous phase change  for large condensates.

 As a final caveat we do not have the homogeneous condensates assumed above and should take the details of their trapping into account.
 The causal length ${\bar\xi}\propto (\rho_0\sqrt{\Gamma_0})^{1/2}$ depends upon density
  and will vary across the trap, but vortices should still form in its center if the condensate is sufficiently large, albeit with
  profile-dependent allometric scaling behavior.   In this regard there are many similarities with the analysis of \cite{zurek3}
  for thermal condensates and we would have to modify our analysis appropriately. This letter is rather aiming for a proof of principle,
  that causality could lead to observable changes of phase accessible by current experiments.

\section*{Acknowledgements}

RR would like to thank the
National Dong Hwa University, Hua-Lien, for support and
hospitality, where much of this work was performed and Dr. Arttu Rajantie, Profs. Randy Hulet and Matt Davis for helpful comments. The work of
DSL and CYL was supported in part by the National Science Council
and the National Center for Theoretical Sciences, Taiwan.

\end{document}